\documentclass[aip,pop,floats,reprint,superscriptaddress]{revtex4-1}
\usepackage{amssymb}
\usepackage{amsmath}
\usepackage{graphicx}

\begin{document}
 \title{Generation of controllable plasma wakefield noise in particle-in-cell simulations}
 \author{N. Moschuering}
 \author{H.Ruhl}
 \affiliation{Ludwig-Maximilians-Universit\"{a}t, 80539 Munich, Germany}
 \author{R.I.Spitsyn}
 \author{K.V.Lotov}
 \affiliation{Budker Institute of Nuclear Physics SB RAS, 630090, Novosibirsk, Russia}
 \affiliation{Novosibirsk State University, 630090, Novosibirsk, Russia}
 \date{\today}
 \begin{abstract}
Numerical simulations of beam-plasma instabilities may produce quantitatively incorrect results because of unrealistically high initial noise from which the instabilities develop. Of particular importance is the wakefield noise, the potential perturbations that have a phase velocity which is equal to the beam velocity. Controlling the noise level in simulations may offer the possibility of extrapolating simulation results to the more realistic low-noise case. We propose a novel method for generating wakefield noise with a controllable amplitude by randomly located charged rods propagating ahead of the beam. We also illustrate the method with particle-in-cell simulations. The generation of this noise is not accompanied by parasitic Cherenkov radiation waves.
 \end{abstract}
 \maketitle

\section{Introduction}

Understanding a collective interaction of relativistic charged particle beams with plasmas is important for a wide variety of physical problems. Among them are space plasmas\cite{PoP17-120501}, fast ignition schemes for inertial fusion\cite{PoP14-055502,PoP12-057305}, turbulent plasma heating for magnetic fusion\cite{FST55(2T)-63,PoP13-062312,PoP17-083111}, positron bunch instabilities driven by an electron cloud in colliders\cite{PRST-AB7-124801}, plasma wakefield acceleration,\cite{PRL104-255003,PoP22-103110,PoP22-123107} and many others. In most cases, the interaction has the form of an instability that develops starting from some low-amplitude shot noise.

A realistic noise level\cite{PRST-AB16-041301} is difficult to reproduce in numerical simulations, since the number of simulated quasi-particles (or macro-particles) is usually much smaller than the number of real particles in the system. Fewer quasi-particles, each carrying a larger charge, produce random fields with amplitudes which are orders of magnitude too high in comparison with the experiment. This, in turn, may result in faster growth of unstable perturbations and quantitatively wrong simulation results.

\begin{figure}[b]
\includegraphics[width=194bp]{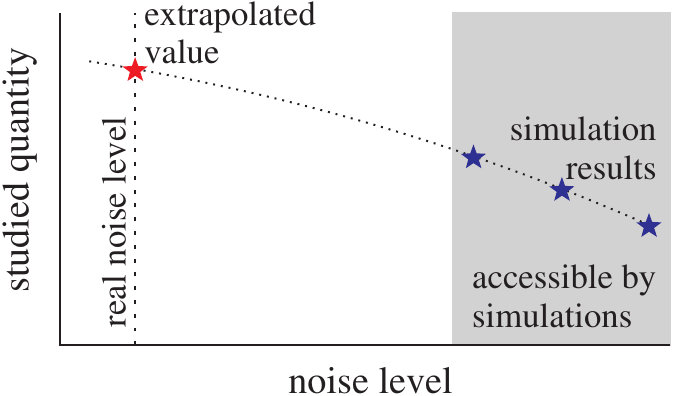}
\caption{Illustration of the extrapolation approach.}\label{fig1-approach}
\end{figure}
Since one-to-one simulations of typical beam-plasma systems of interest fall far beyond state-of-the-art computational capabilities, the only available choices are ordered initial distributions of particles\cite{PRST-AB16-041301} and extrapolating high-noise simulation results to low-noise physical systems. The second approach is schematically illustrated in Fig.\,\ref{fig1-approach}. Obviously it would benefit from having several simulation points with different noise levels. This makes it desirable to develop a simple method with the capability to produce a noise field with a controllable amplitude. For reliable extrapolation, the noise level must be controlled independently of key simulation parameters, like the grid resolution or the number of quasi-particles.

\begin{figure}[b]
\includegraphics[width=219bp]{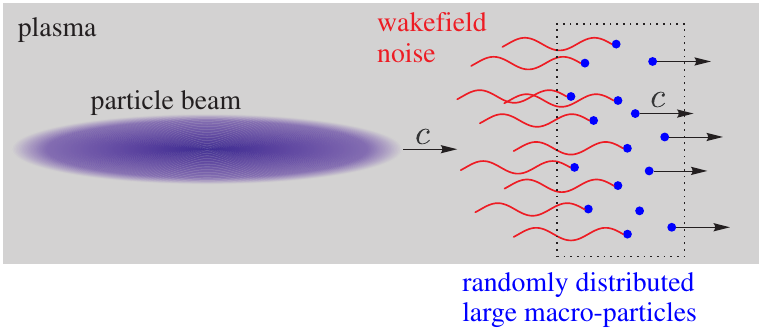}
\caption{Illustration of the controllable noise generation.}\label{fig2-idea}
\end{figure}
For many instabilities of relativistic beams, the noise harmonics of interest are potential (Langmuir) plasma waves with a phase velocity which is close to the speed of light $c$ and the wavevector $\vec{k}$ directed along the axis of beam propagation. We will call these waves wakefield noise, which is an analogy to regularly excited plasma wakefields, which have the same properties. At first glance, a controllable wakefield noise might be easily excited by an ensemble of randomly located point-like quasi-particles propagating ahead of the beam (Fig.\,\ref{fig2-idea}). By changing the charge of the quasi-particles and their number, it is possible to control the noise level. However, relativistic quasi-particles usually emit numerical Cherenkov radiation, if simulated using particle-in-cell codes.\cite{JCP15-504,PRST-AB16-021301,EPJD68-177,NIMA-829-353} This radiation critically affects a clean study of the beam instability and must therefore be avoided.

In this paper we propose a novel method for generating wakefield noise with a controllable amplitude by randomly located charged rods (Sec.\,\ref{s2}). The noise is free from parasitic Cherenkov radiation. We also give expressions for the relations between the amplitude of the noise field and the rod parameters (Sec.\,\ref{s3}) and illustrate the method using particle-in-cell (PIC) simulations (Sec.\,\ref{s4}). The main findings are summarized in Sec.\,\ref{s5}. Studies of particular beam instabilities fall beyond the scope of this paper.

For all simulations detailed in this paper we use the PIC code PSC.\cite{PSC} The coordinates are either Cartesian $(x,y,z)$ or cylindrical $(r, \phi, z)$ with the $z$-axis being the direction of beam propagation. The co-moving coordinate $\xi=z-ct$ is used wherever convenient.

\section{The idea of charged rods}
\label{s2}

\begin{figure}[t]
\includegraphics[width=230bp]{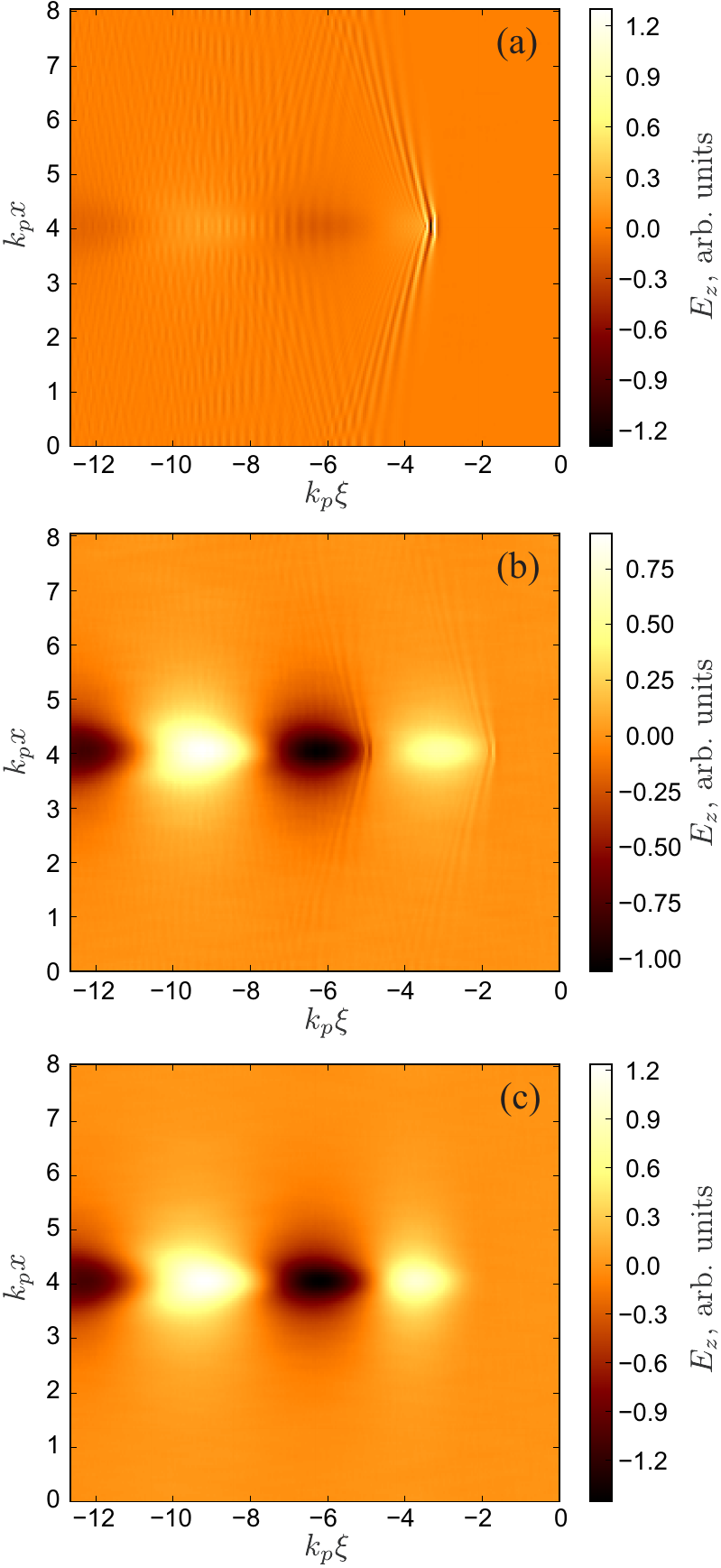}
\caption{Simulated wake patterns for moving objects of different shapes: point-like charge (a), rectangular-shaped rod (b), and cosine-shaped rod (c).}\label{fig3-cherenkov}
\end{figure}

The problem of numerical Cherenkov radiation depends on the details of the numerical solver. In the present paper we use of the finite-difference time-domain (FDTD) scheme. The discretization of the FDTD scheme leads to a phase velocity of the propagated radiation, which is strictly smaller than the velocity of light. A quasi-particle with sufficient energy can therefore travel faster than the radiation on the grid. This leads to numerical Cherenkov radiation, as illustrated in Fig.\,\ref{fig3-cherenkov}(a). A point-like charge $Q$ moving with the speed of light emits short-wavelength electromagnetic radiation, the amplitude of which is much higher than the amplitude of the useful plasma wave. The radiation wavelength is of the order of the grid size and is much shorter than the plasma wavelength $\lambda_p = 2\pi k_p^{-1}$. The plasma wavelength is determined by the plasma density $n_0$ through the plasma wavenumber $k_p = \sqrt{4 \pi n_0 e^2/ (m c^2)}$, where $e$ is the elementary charge and $m$ is the electron mass. The amplitude of the emitted Cherenkov electromagnetic waves is proportional to the amplitude of short-wavelength harmonics in the Fourier spectrum of the emitting charge. More specifically, the $z$-component of the wavevector is of importance. This already hints at the idea of how the numerical Cherenkov radiation can be reduced in comparison to the longer-wavelength plasma waves: The radiating source must be long and smooth. In other words, it should be shaped like a charged rod.

We consider two variants of rods: i) Rectangular rods with the linear charge density
\begin{equation}\label{e1}
    \lambda(\xi) = k_p Q/\pi, \qquad -\lambda_p/2 < \xi-\xi_0 < 0,
\end{equation}
and ii) Cosine-shaped rods with
\begin{equation}\label{e2}
    \lambda(\xi) = \frac{k_p Q}{\pi} \Bigl(1-\cos\bigl(2 k_p (\xi-\xi_0)\bigr) \Bigr), \quad -\lambda_p/2 < \xi-\xi_0 < 0.
\end{equation}
For efficient excitation of Langmuir waves, the rod length must be $\lessapprox \lambda_p$. We set it to $\lambda_p/2$. This value does not maximize the wakefield of the rod and is chosen only for the simplification of the subsequent analytical calculations it provides.

\begin{figure}[t]
\includegraphics[width=209bp]{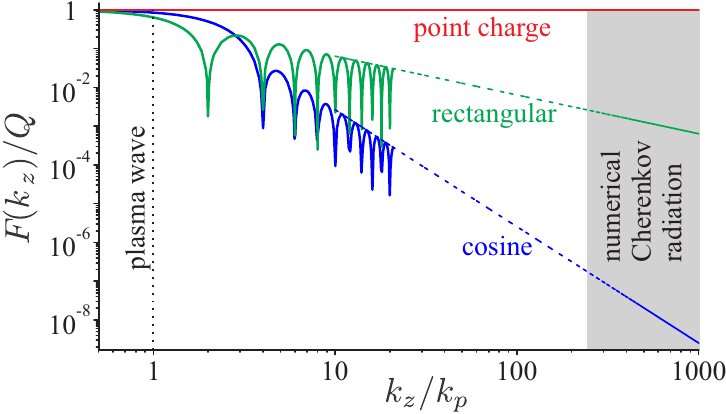}
\caption{Fourier spectra of various charge distributions. For large $k_z$, only the envelopes of the oscillating spectra are shown.}\label{fig4-spectra}
\end{figure}

The Fourier spectra
\begin{equation}\label{e3}
    F(k_z) = \left| \int_{-\infty}^\infty \lambda(\xi) e^{i k_z \xi} d\xi \right|
\end{equation}
of these rods and for the point charge $Q$ are shown in Fig.\,\ref{fig4-spectra}. For the wave number of plasma waves holds that $k_z \approx k_p$ (vertical dotted line in Fig.\,\ref{fig4-spectra}) and they are excited equally well by all charge distributions. Numerical Cherenkov waves for typical resolutions of PIC code simulations have wave numbers for which $k_z/k_p \sim 10^2 - 10^3$ (shaded area in Fig.\,\ref{fig4-spectra}) holds, and their excitation is suppressed by some orders of magnitude in the case of cosine-like rods. The smoothness degree of the charge distribution $\lambda(\xi)$ determines the decrease rate of the spectrum $F(k_z)$ for high $k_z$. In the case of the cosine-like distribution this decrease rate is given by $F(k_z) \propto k_z^{-3}$.

A ``beam'' of equally charged rods would produce not only the noise field, but also a regular wakefield which is excited by the total rod charge. This is undesirable. To get rid of the regular field, the sign of the charge of the different rods must be chosen randomly.

Another necessary condition for achieving a correct noise is the equivalence of all wakefield phases. The contributions of different rods must, on average, uniformly cover the whole period of the plasma wave. Using alternating rod charges the minimum length of the rod area is $\lambda_p/2$, as rods of opposite charge contribute to opposite half-periods of the wave.

Note that an exact nulling of the total charge of the rod ensemble, for example by choosing an equal number of positively and negatively charged rods, is erroneous. This additional constraint on the rod ensemble would break the equivalence of wakefield phases. This can be easily seen in the extreme case of two oppositely charged rods.

\section{Amplitude of the wakefield noise}
\label{s3}

In this section we will analytically calculate the root mean square (rms) of the longitudinal electric field excited by an ensemble of cosine-shaped rods \eqref{e2}. Assume that the rod heads (characterized by random coordinates $\vec{r}_{\perp 0}$ and $\xi_0$) are uniformly distributed in $z$-direction over an infinitely wide layer of thickness $\pi k_p^{-1}$. The average density in this region is given by $2n$, with $n$ being the average number density of rods of each charge sign ($Q$ or $-Q$).

We will first find an expression for the wakefield of a single rod. Given the charge density of the rod
\begin{equation}\label{e4}
    \rho(\vec{r}_\perp,\xi) = \delta (\vec{r}_\perp) \lambda(\xi), \qquad \vec{r}_\perp = (x,y),
\end{equation}
the longitudinal component of the wakefield is\cite{PAcc22-81}
\begin{multline}\label{ez_gen}
    E_z(\vec{r}_\perp,\xi) = 2{k_p}^2 \int_{\xi}^{\infty} d\xi' \int d \vec{r}_\perp{\!\!\!}'\; \rho(\vec{r}_\perp{\!\!\!}'\:,\xi')\\
    \times K_0 \left(k_p|\vec{r}_\perp-\vec{r}_\perp{\!\!\!}'\:|\right) \cos\bigl(k_p(\xi-\xi')\bigr) \\
     = \frac{2 k_p^2 Q}{\pi} K_0 (k_p r) G(\xi-\xi_0),
\end{multline}
where $K_0$ is the modified Bessel function of the second kind, and
\begin{multline}
G(\xi) = k_p \int_{\max(-\pi k_p^{-1}, \xi)}^{0} d\xi' \bigl(1-\cos(2 k_p \xi') \bigr) \cos\bigl(k_p(\xi-\xi')\bigr) \\
= \begin{cases}
   -\frac{8}{3}\sin(\xi), & k_p \xi<-\pi,\\
   \frac{2}{3}\bigl(\sin(2\xi) - 2\sin(\xi)\bigr), &-\pi<k_p \xi<0,\\
   0, &\xi>0.
 \end{cases}
\end{multline}

In Ref.\,\onlinecite{PRST-AB16-041301}, the rms noise field of $N$ charges is calculated as
\begin{equation}
E_\text{rms}^2=N\langle {E_z}^2 \rangle,
\end{equation}
where the angle brackets denote the averaging of the field of a single charge over possible locations of this charge with respect to the observation point. We have to modify our approach in this case, as we take the number of rods to be infinite, which makes the average field vanish. Assume the rods occupy a cylindrical area with a large radius $R$. Then
\begin{equation}\label{e7}
    N=2\pi^2 k_p^{-1} R^2 n
\end{equation}
and the average field
\begin{multline}\label{e9}
    E_\text{rms}^2 (\xi) = N \left\langle\left(\frac{2 k_p^2 Q}{\pi} K_0 (k_p r) G(\xi-\xi_0)\right)^2\right\rangle \\
    = 2n \frac{4 k_p^4 Q^2}{\pi^2} \int_0^R 2 \pi r K_0^2 (k_p r) dr \,
        \int_\xi^0 G^2(\xi-\xi_0) \, d\xi_0 \\
    = \frac{8 k_p^2 Q^2 n}{\pi} \Bigl(1+k_p^2 R^2 \bigl(K_0^2(k_p R) - K_1^2(k_p R)\bigr)\Bigr) \\
    \times \int_\xi^0 G^2(\xi-\xi_0) \, d\xi_0.
\end{multline}
In the limit $R \to \infty$ it holds that
\begin{equation}\label{e10}
    E_\text{rms}^2 (\xi) = \frac{8 k_p^2 Q^2 n}{\pi} \int_\xi^0 G^2(\xi-\xi_0) \, d\xi_0,
\end{equation}
and for $\xi < -\pi k_p^{-1}$ (trailing the rods),
\begin{equation}\label{e10a}
    E_\text{rms}^2 (-\infty) = \frac{256 k_p Q^2 n}{9}.
\end{equation}

\section{Wakefield noise in PIC simulations}
\label{s4}

We now describe our approach to the production of wakefield noise in three-dimensional PIC simulations, how the noise looks like, and which additional actions are required to observe a close quantitative agreement between simulation results and the developed theory.

We configure the code to use widespread algorithms: a standard Boris pusher\cite{Boris} with second order particles to perform particle pushing and a standard FDTD-scheme\cite{FDTD} to perform field pushing. The simulation window moves with the speed of light in a patch-based manner, that is, by appending a simulation grid and quasi-particles on one end of the box and detaching the same amount of volume on the opposite side. This process is iterated continuously. The window size is given by $X \times Y \times Z$ with $X=Y=0.3 \lambda_p$, $Z=2.02 \lambda_p$. The boundary conditions are periodic in transverse dimensions ($x$ and $y$) and reflecting in the $z$ dimension. The simulated propagation length, moving in the $z$ direction, is $10 \lambda_p$.
Initially the plasma is cold and uniform. It is composed using 3 quasi-particles per cell for the electrons, while the ions are not simulated and treated as immobile charges. The spatial resolutions are $\Delta x = \Delta y = \Delta z = \lambda_p / 130 \approx 0.05 k_p^{-1}$ and the time step is $\Delta t = 0.99 \Delta x / (c\sqrt{3}) \approx 0.28 \omega_p^{-1}$. The simulated time for the whole simulation is given by $T = 10\lambda_p/c$, which results in a number of time steps of about $2260$.

At the start of the simulation, $100$ random positions are computed in the cuboid given by $0 \le x < X$, $0 \le y < Y$ and $1.5 \lambda_p \le z \le 2 \lambda_p$. These positions $\left(x_i, y_i, z_i\right)$ are used as head positions for $100$ rods of length $\lambda_p/2$. The rods are constructed as strings of quasi-particles with a distance of $\Delta z$ between each adjacent pair. The strings start at $\left(x_i, y_i, z_i - 0.5\Delta z\right)$ and continue with decreasing z-coordinates. These quasi-particles have the same shape as the plasma quasi-particles, a relativistic factor of $\gamma = 10^{10}$, and a charge that varies according to the cosine-like distribution \eqref{e2}. The total charge of a single rod (the sum of all charges in the constituting string of quasi-particles) is $|Q| \approx 1.8 \times 10^{-3} e n_0 k_p^{-3}$, where the sign of the charge is chosen randomly. This charge is sufficiently high for a clear observation of the wakefield and sufficiently low to stay within the regime of a linear plasma response.

To enhance the wakefield noise against other noise harmonics (which are always present in PIC simulations), we average the fields over many time steps according to the formula
\begin{equation}\label{e11}
    \bar{E}_z\left(x, y, \xi\right) = \frac{1}{1000}\sum_{t=1201 \Delta t}^{2200 \Delta t} E_z\left(x, y, \xi, t\right).
\end{equation}
This averaging strongly suppresses all perturbations except those propagating with the speed of light and being stationary in the co-moving frame.

\begin{figure}[t]
\includegraphics[width=236bp]{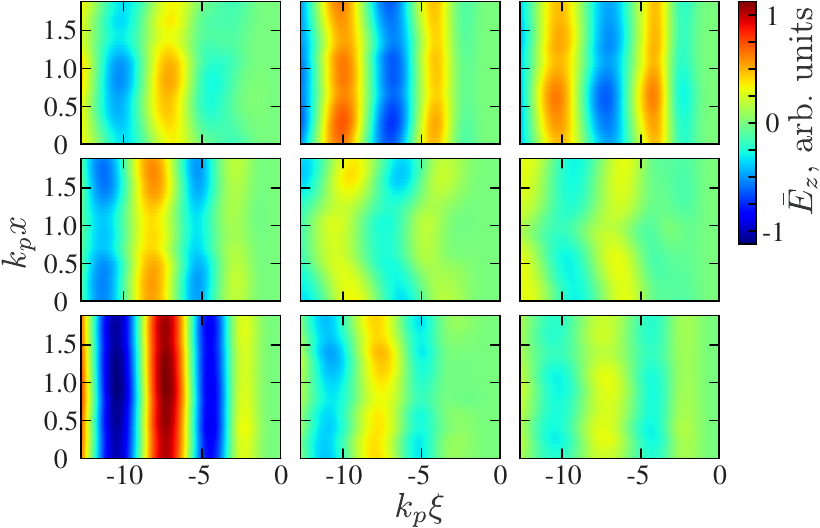}
\caption{The longitudinal wakefield $\bar{E}_z (x, y, \xi)$ at $y=0.15 \lambda_p$ for several rod distributions.}\label{fig5-maps}
\end{figure}

\begin{figure}[b]
\includegraphics[width=226bp]{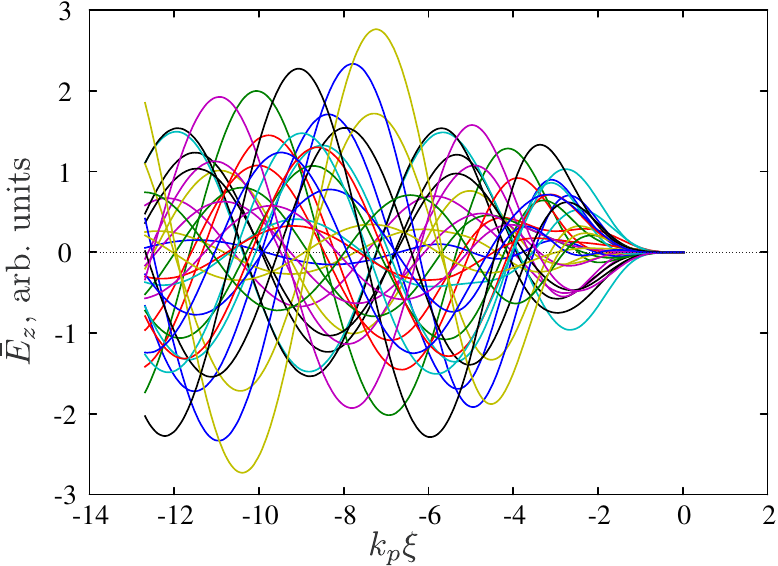}
\caption{The longitudinal wakefield $\bar{E}_z (\xi)$ at $x=y=0.15 \lambda_p$ generated by 100 cosine rods for 32 different random seeds.}\label{fig6-many}
\end{figure}

The simulation is then run for a large number of samples $n_s$ with different random number generator seeds, which leads to different rod positions. Examples of the produced field distributions are shown in Fig.\,\ref{fig5-maps}. Since the frequency $\omega_p$ and the phase velocity $c$ of the perturbation are fixed, the wavelength $\lambda_p$ is fixed as well, which makes the wakefield noise periodic in $\xi$. It therefore does not look like usual noise. The transverse distance for field correlations is about $k_p^{-1}$, which makes random field changes in transverse directions only visible in the case of wider simulation areas. The amplitude and the phase of the generated fields are quite random even for narrow simulation areas (Fig.\,\ref{fig6-many}), and their properties can be characterized by the mean value $\mu(\xi)$ and the standard deviation $\sigma(\xi)$:
\begin{equation}\label{e12}
    \mu(\xi) = \frac{1}{n_s}\sum_{i=1}^{n_s}\bar{E}_z^i\left(x_m, y_m, \xi\right),
\end{equation}
\begin{equation}\label{e13}
\sigma(\xi)=\sqrt{\frac{1}{n_s-1} \sum_{i=1}^{n_s}\left(\bar{E}_z^i\left(x_m, y_m, \xi\right) - \mu\left(\xi\right)\right)^2},
\end{equation}
where the superscript $i$ denotes the sample number, and $x_m$ and $y_m$ are the coordinates of the observation line. For the following we put $x_m=y_m=0.15 \lambda_p$.

\begin{figure}[tb]
\includegraphics[width=215bp]{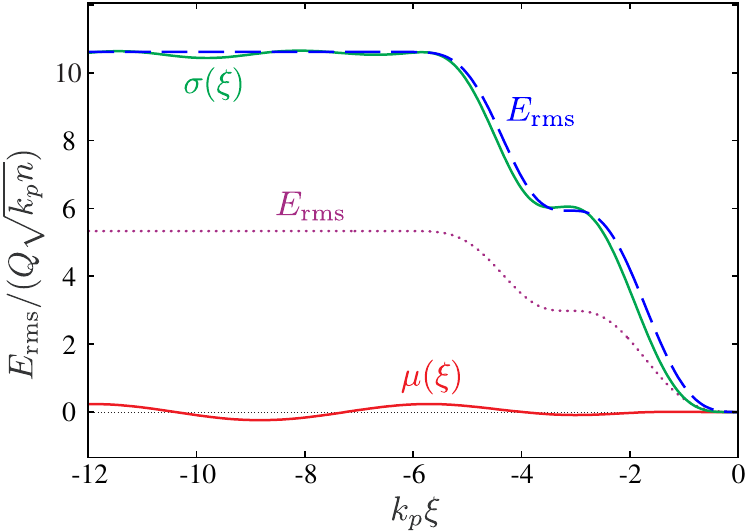}
\caption{The mean $\mu(\xi)$ and the standard deviation $\sigma(\xi)$ of the simulated wakefield (using 11264 rod samples with 100 rods each) and the calculated average $E_\text{rms} (\xi)$ with (dashed line) and without (dotted line) taking into account correlation effects due to the periodical boundary conditions.}\label{fig7-average}
\end{figure}
\begin{figure}[tb]
\includegraphics[width=226bp]{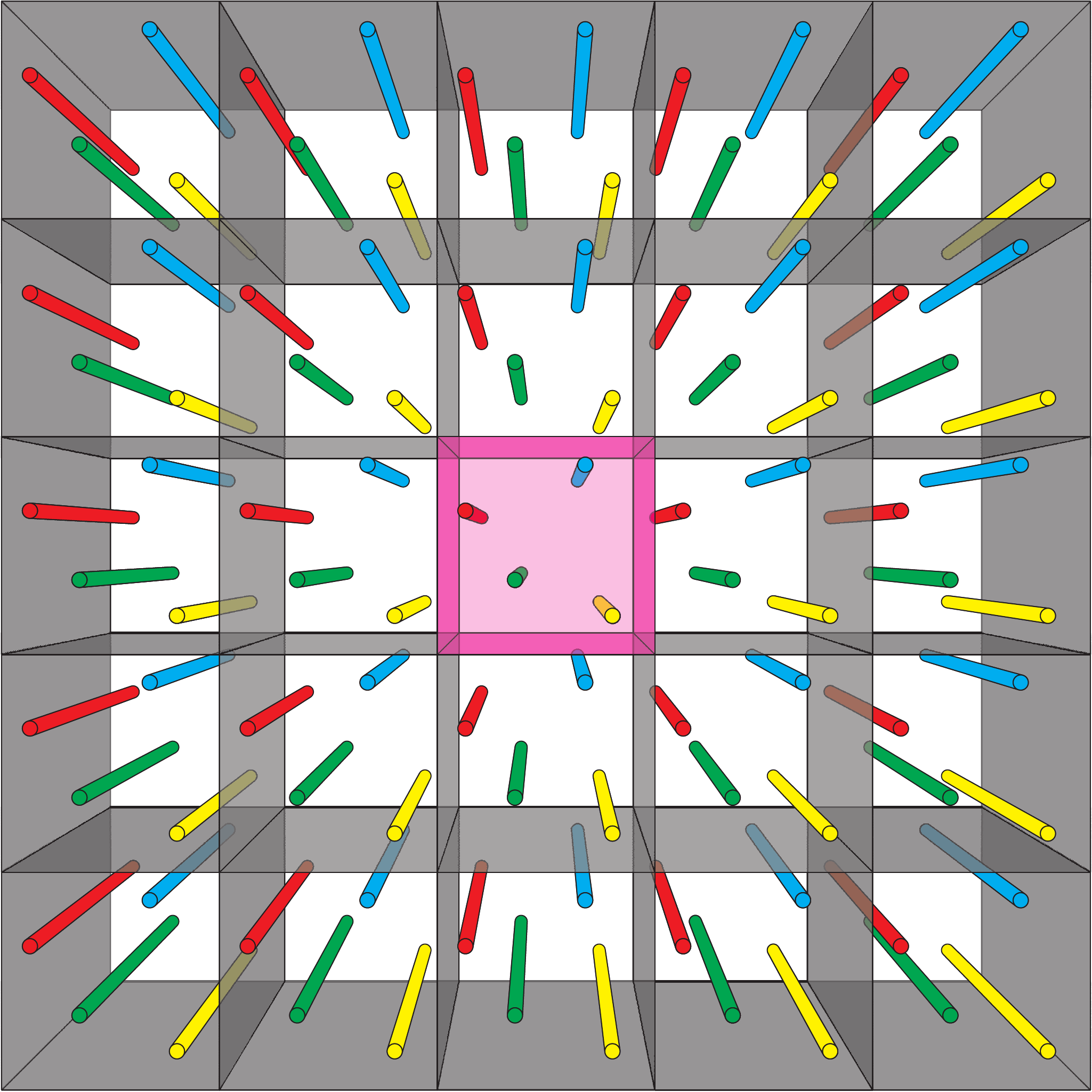}
\caption{Illustration of the effects of periodic boundaries. Physical processes in the rectangular area (shown in pink) simulated with periodic boundaries are equivalent to processes with a domain where the entire, infinite space is composed of identical replications of the pink area (replicas shown by the boxes with grey walls). Rods of the same color are identical in charge and longitudinal position and are located strictly periodically in transverse directions.}\label{fig8-boundary}
\end{figure}

The resulting values for $\mu(\xi)$ and $\sigma(\xi)$ are shown in Fig.\,\ref{fig7-average}. The dotted line in Fig.\,\ref{fig7-average} gives the value of $E_\text{rms}$ calculated for these rods according to the formula \eqref{e10}. It is substantially lower than the established average field in simulations. The difference results from the periodical boundary conditions. Each rod in the simulation domain has an infinite number of replicas, the positions of which are correlated with the rod position (Fig.\,\ref{fig8-boundary}). Consequently, for correct comparison of the theory and the simulation results, the radial averaging in \eqref{e9} has to be modified:
\begin{multline}\label{e14}
    \int_0^R 2 \pi r K_0^2 (k_p r) dr \ \to \\  \
    \int_{-X/2}^{X/2} \int_{-Y/2}^{Y/2} dx \, dy \, \sum_{i,j} K_0^2 (k_p r_{ij}),
\end{multline}
where
\begin{equation}\label{e15}
    r_{ij} = \left| \vec{r}_\perp + i X \vec{e}_x + j Y \vec{e}_y \right|,
\end{equation}
$\vec{e}_x$ and $\vec{e}_y$ are unit vectors, and indices $i,j \in \mathbb{Z}$, where $i=j=0$ corresponds to the position of one specific rod (situated in the pink area in Fig.\,\ref{fig8-boundary}) and all other values correspond to the positions of its replicas (situated in gray boxes in Fig.\,\ref{fig8-boundary}). The dashed line in Fig.\,\ref{fig7-average} shows the modified values for the average field.

\section{Summary}
\label{s5}

We have shown a reliable method of producing controllable noise levels in PIC simulations. This controllable wakefield noise does not suffer from numerical Cherenkov radiation. Analytical expressions for the rms amplitude of this noise have been given. These expressions include formulas for two different external boundaries, namely for periodic boundaries as well as for open or absorbing boundaries. These expressions and the method in general were illustrated using very common PIC simulation techniques, making them very comparable and applicable for the research community. A very good agreement has been found. This method can be used to perform noise level scans for a multitude of noise seeded physical processes. This paper also provides an understanding of the to-be-expected noise structure. Using these noise level scans, the correct behavior can be predicted by extrapolation of the generated data points. Further research should focus on the application of this novel method on specific beam-plasma non-linearities.

\acknowledgments

Contribution of R.S. and K.L. to this work is supported by The Russian Science Foundation, grant No.~14-50-00080. N. M. and H. R. acknowledge the hospitality of the Arnold Sommerfeld Center for Theoretical Physics at the Ludwig Maximilians University. N. M. acknowledges the very useful advice of K. U. Bamberg during the development of the computational simulation. This work was supported by the Cluster-of-Excellence Munich Centre for Advanced Photonics (MAP) and the Gauss Centre for Supercomputing (GCS), project PLASMA SIMULATION CODE, LRZ-ID pr84me.


\begin{thebibliography}{88}
 \bibitem{PoP17-120501}
    A. Bret, L. Gremillet, and M. E. Dieckmann,
    Phys. Plasmas \textbf{17}, 120501 (2010).
 \bibitem{PoP14-055502}
    M. H. Key,
    Phys. Plasmas \textbf{14}, 055502 (2007).
 \bibitem{PoP12-057305}
    M. Tabak, D. S. Clark, S. P. Hatchett, M. H. Key, B. F. Lasinski, R. A. Snavely, S. C. Wilks, R. P. J. Town, R. Stephens, E. M. Campbell, R. Kodama, K. Mima, K. A. Tanaka, S. Atzeni, R. Freeman,
    Phys. Plasmas \textbf{12}, 057305 (2005).
\bibitem{FST55(2T)-63}
    A. Burdakov, A. Arzhannikov, V. Astrelin, V. Batkin, V. Burmasov, G. Derevyankin, V. Ivanenko, I. Ivanov, M. Ivantsivskiy, I. Kandaurov, V. Konyukhov, K. Kuklin, S. Kuznetsov, A. Makarov, M. Makarov, K. Mekler, S. Polosatkin, S. Popov, V. Postupaev, A. Rovenskikh, A. Shoshin, S. Sinitsky, V. Stepanov, Yu. Sulyaev, Yu. Trunev, L. Vyacheslavov, Ed. Zubairov,
    Fusion Sci. Technol. \textbf{55}(2T), 63 (2009).
 \bibitem{PoP13-062312}
    I. V. Timofeev and K. V. Lotov,
    Phys. Plasmas \textbf{13}, 062312 (2006).
 \bibitem{PoP17-083111}
    I. V. Timofeev and A. V. Terekhov,
    Phys. Plasmas \textbf{17}, 083111 (2010).
 \bibitem{PRST-AB7-124801}
    F. Zimmermann,
    Phys. Rev. ST  Accel. Beams \textbf{7}, 124801 (2004).
\bibitem{PRL104-255003}
	N. Kumar, A. Pukhov, and K. Lotov,
	Phys. Rev. Lett. \textbf{104}, 255003 (2010).
\bibitem{PoP22-103110}
	K. V. Lotov,
	Phys. Plasmas \textbf{22}, 103110 (2015).
\bibitem{PoP22-123107}
    K. V. Lotov,
    Phys. Plasmas \textbf{22}, 123107 (2015).
 \bibitem{PRST-AB16-041301}
	K. V. Lotov, G. Z. Lotova, V. I. Lotov, A. Upadhyay, T. Tuckmantel, A. Pukhov, A. Caldwell,
	Phys. Rev. ST Accel. Beams \textbf{16}, 041301 (2013).
\bibitem{JCP15-504}
    B. N. Godfrey,
    Journal of Computational Physics \textbf{15}, 504 (1974).
\bibitem{PRST-AB16-021301}
    R. Lehe, A. Lifschitz, C. Thaury, V. Malka, X. Davoine,
    Phys. Rev. ST Accel. Beams \textbf{16}, 021301 (2013).
\bibitem{EPJD68-177}
    R. Nuter, M. Grech, P. Gonzalez de Alaiza Martinez, G. Bonnaud, E. d'Humieres,
    Eur. Phys. J. D \textbf{68}, 177 (2014).
\bibitem{NIMA-829-353}
    J.-L. Vay, R. Lehe, H. Vincenti, B. B. Godfrey, I. Haber, P. Lee,
    Nuclear Instr. Meth. A \textbf{829}, 353 (2016).
\bibitem{PSC}
    K. Germaschewski, W. Fox, S. Abbott, N. Ahmadi, K. Maynard, L. Wang, H. Ruhl, A. Bhattacharjee,
    Journal of Computational Physics \textbf{318}, 305 (2016).
  \bibitem{PAcc22-81}
    T. Katsouleas, S. Wilks, P. Chen, J. M. Dawson, and J. J. Su,
    Part.Accel., \textbf{22}, 81 (1987).
\bibitem{Boris}
    J. P. Boris,
    Proceedings of the 4th Conference on Numerical Simulation of Plasmas. Naval Res. Lab., Washington, D.C., 3--67 (1970).
\bibitem{FDTD}
    K. Yee,
    IEEE Transactions on Antennas and Propagation \textbf{14(3)}, 302 (1966).

\end{thebibliography}
\end{document}